\documentclass[reprint,aps,twocolumn,superscriptaddress,nofootinbib]{revtex4-2}

\usepackage{bm}% bold math
\usepackage{dcolumn}% Align table columns on decimal point
\newcommand{\bra}[1]{\langle{#1}|}
\newcommand{\ket}[1]{|{#1}\rangle}

\newcommand{\tonde}[1]{\left({#1}\right)}

\newcommand{\graffe}[1]{\left\{{#1}\right\}}
\usepackage{graphicx}
\usepackage{tabularx}
\usepackage{nicefrac}
\usepackage{amsmath}

\begin{document}

\preprint{APS/123-QED}

\title{Generation of hyperentangled photon pairs\\
in the time and frequency domain on a silicon photonic chip}

\author{Sara Congia}
 \email{sara.congia02@universitadipavia.it}
\affiliation{Dipartimento di Fisica "Alessandro Volta", Università di Pavia, via Bassi 6, 27100 Pavia, Italy}
\affiliation{Université Grenoble Alpes, CEA-Leti, 38054 Grenoble, France}

\author{Massimo Borghi}
\affiliation{Dipartimento di Fisica "Alessandro Volta", Università di Pavia, via Bassi 6, 27100 Pavia, Italy}

\author{Emanuele Brusaschi}
\affiliation{Dipartimento di Fisica "Alessandro Volta", Università di Pavia, via Bassi 6, 27100 Pavia, Italy}

\author{Federico Andrea Sabattoli}
\altaffiliation[Current address: ]{Advanced Fiber Resources Milan S.r.L, via Fellini 4, 20097 San Donato Milanese, Italy;}
\affiliation{Dipartimento di Fisica "Alessandro Volta", Università di Pavia, via Bassi 6, 27100 Pavia, Italy}

\author{Houssein El Dirani}
\altaffiliation[Current address: ]{LIGENTEC SA, 224 Bd John Kennedy, 91100 Corbeil-Essonnes, France;}
\affiliation{Université Grenoble Alpes, CEA-Leti, 38054 Grenoble, France}

\author{Laurene Youssef}
\altaffiliation[Current address: ]{ENSIL-ENSCI/IRCER, UMR CNRS 7315, Université de Limoges, France;}
\affiliation{Université Grenoble Alpes, CEA-Leti, 38054 Grenoble, France}

\author{Camille Petit-Etienne}
\altaffiliation[Current address: ]{Univ. Grenoble Alpes, CNRS, CEA/LETI-Minatec, Grenoble INP, LTM, Grenoble F-38054, France;}
\affiliation{Université Grenoble Alpes, CEA-Leti, 38054 Grenoble, France}

\author{Erwine Pargon}
\altaffiliation[Current address: ]{Univ. Grenoble Alpes, CNRS, CEA/LETI-Minatec, Grenoble INP, LTM, Grenoble F-38054, France;}
\affiliation{Université Grenoble Alpes, CEA-Leti, 38054 Grenoble, France}

\author{Corrado Sciancalepore}
\altaffiliation[Current address: ]{STMicroelectronics, 38920 Crolles, France}
\affiliation{Université Grenoble Alpes, CEA-Leti, 38054 Grenoble, France}

\author{Marco Liscidini}
\affiliation{Dipartimento di Fisica "Alessandro Volta", Università di Pavia, via Bassi 6, 27100 Pavia, Italy}

\author{Johan Rothman}
\affiliation{Université Grenoble Alpes, CEA-Leti, 38054 Grenoble, France}

\author{Ségolène Olivier}
\affiliation{Université Grenoble Alpes, CEA-Leti, 38054 Grenoble, France}

\author{Matteo Galli}
\affiliation{Dipartimento di Fisica "Alessandro Volta", Università di Pavia, via Bassi 6, 27100 Pavia, Italy}

\author{Daniele Bajoni}
\affiliation{Dipartimento di Ingegneria Industriale e dell’Informazione, Università di Pavia, via Ferrata 1, 27100 Pavia, Italy}

\date{\today}% It is always \today, today,
             %  but any date may be explicitly specified

\begin{abstract}
Multi-dimensional entangled photon states represent an important resource in quantum communication networks. Specifically, hyperentangled states presenting simultaneous entanglement in several degrees of freedom (DoF), stand out for their noise resilience and information capacity. In this work, we demonstrate the generation of hyperentangled photon pairs in the time and frequency-bin domain by spontaneous four-wave mixing from the coherent driving of two integrated Silicon microresonators. We demonstrate entanglement in each DoF by proving the violation of the Clauser Horne Shimony Holt (CHSH) inequality by more than 27 standard deviations (STDs) in each reduced space. Genuine hyperentanglement is then assessed from the negativity of an hyperentanglement witness, which is verified by more than 60 STDs. These results mark, to the best of our knowledge, the first demonstration of time-frequency bin hyperentanglement in an integrated silicon photonic device.
\end{abstract}

%\keywords{Suggested keywords}%Use showkeys class option if keyword
                              %display desired
\maketitle

\section{Introduction}
The simultaneous entanglement of photons in multiple degrees of freedom, namely hyperentanglement (HE) \cite{1997_Kwiat_HE}, offers several unique capabilities, from the realization of Bell-state analyzers using linear optics \cite{2006_Schuck_HBSA, 2007_Wei_HBSA}, super-dense coding \cite{2008_Barreiro_superdence, 2017_williams_superdense} and cluster state quantum computation \cite{2007_Chen_HEtoCluster, 2008_Vallone_HEtoCluster, 2019_Morandotti_HEtimefreq}. Additionally, HE states can be directly employed in quantum communication protocols, showing higher robustness against decoherence in lossy channels with respect to multi-photon quantum states with equal Hilbert space dimensionality \cite{2021_Kim_noiseresistant}, or enabling high-fidelity entanglement distillation \cite{2023_Xu_Distillation_PolFreq}.\\
\indent To date, most of the experimental demonstrations of HE rely on photon pairs generated by spontaneous parametric down conversion (SPDC) in bulk non-linear crystals \cite{2017_Deng_HE}. Within this framework, HE photon pairs have been generated in polarization, momentum, and energy-time DoF \cite{2005_Barreiro_3DoFs, 2009_Ceccarelli_Vallone_3DoFs, 2009_Vallone_Ceccarelli_3DoFs, 2023_Lu_ultrabroad},\ reaching space-dimensions equivalent to five \cite{2015_Xie_5qubit} and ten qubits \cite{2010_Gao_Cat}. Increasing efforts have been dedicated to miniaturizing photon-pair sources and integrating them on quantum photonic chips, bringing benefits in terms of scalability, compactness, stability, and power consumption \cite{2021_Chen_review}. Polarization-time hyperentanglement has been demonstrated with \emph{integrated} sources of photon pairs based on SPDC in periodically poled lithium niobate waveguides \cite{2020_Kumar_LiNbO3, 2022_Huang_LiNbO3}. Spontaneous four wave mixing (SFWM) has been used to generate HE photon pairs in the time-polarization DoF using a Silicon (Si) microresonator \cite{2015_Suo_HEpolatime}, and in the frequency-time bin DoF in a Si-oxynitride resonator \cite{2019_Morandotti_HEtimefreq}. In particular, the time-bin (TB) and frequency-bin (FB) encoding is relevant for its compatibility with \emph{both} fiber links and free space channels.\\
\indent For example, time-bin encoding is suitable for long-distance transmission through fibers \cite{2004_Marcikic_TB50km}, free space \cite{2007_Ursin_144km}, and even for space applications, like ground-satellites links \cite{2016_Vallone_space}, with recent and remarkable demonstrations of the multidimensional entanglement distribution \cite{2020_Vagniluca_TB4, 2025_Morendotti_TB8}. Frequency-bin (FB) encoding can be naturally generated by SFWM in microresonators \cite{2017_Kues_FB, 2023_Clementi, 2023_Borghi} over a high-dimensional space \cite{lu2023frequency}, is robust to logical-errors during fiber-optic transmission \cite{2025_Kues_FB_QKD,2024_Tagliavacche}, and allows gate parallelization over a single spatial mode \cite{henry2023parallelizable}.\\
\indent In this work, we present an integrated source of HE photon pairs over discrete \emph{time-frequency} modes composed by two Si-microrings. The use of multiple resonators allows one to have a simultaneous high-brightness and a small frequency-bin spacing \cite{2019_Liscidini_FBtheory,2023_Clementi, 2023_Borghi}, which are a key requisites for efficient frequency-bin manipulation using commercial-grade electro-optic modulators ($<40\,$GHz electro-optic bandwidth) and high-rate transmission over lossy links. This circumvents the limitation of using a single and large resonator, in which a small frequency-bin spacing is achieved at the cost of a reduced photon pair generation rate \cite{2019_Liscidini_FBtheory}. 
We first performed quantum state tomography of the reduced density matrices over the TB and FB modes, proving the entanglement of the two photons in each DoF. We then used an HE witness \cite{2008_Vallone_witness} to assess that the complete state is entangled. 

\section{Device and experimental setup}
\label{sec:exp_setup}
\begin{figure*}[ht]
\centering
\includegraphics[width = \textwidth]{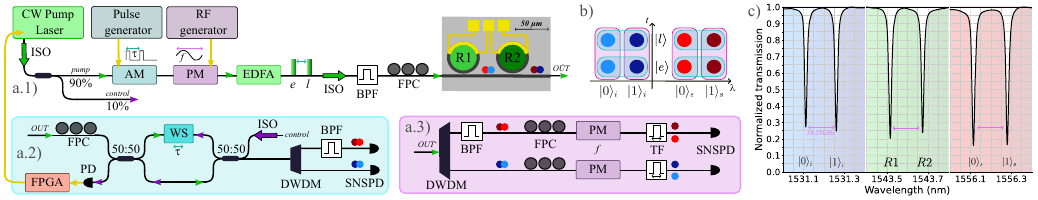}
\caption{\textbf{(a)} Simplified sketch of the experimental setup, divided into its three parts: \textbf{(a.1)} generation setup, \textbf{(a.2)} Time-analyzer (T-a), \textbf{(a.3)} Frequency-analyzer (F-a). The direction of the pump laser and of the generated photons is indicated with a green arrow. The purple arrow indicates the direction of the cw control beam. AM: Amplitude Modulator, PM: Phase Modulator, EDFA: Erbium-Doped Fiber Amplifier, ISO: fiber isolator, BPF: Band-Pass Filter, FPC: Fiber Polarization Controller, WS: WaveShaper, PD: PhotoDiode, FPGA: Field Programmable Gate Array, DWDM: Dense Wavelength Division Multiplexing, TF: Tunable Filter. \textbf{(b)} Schematic representation of the target two-photon HE state; the modes are defined as $\ket{0}$ and $\ket{1}$ for the FB and $\ket{e}$ and $\ket{l}$ for the TB DoF, respectively. \textbf{(c)} Transmission spectra in the ranges on interest, i.e. from left to right: idler, pump and signal resonances.}
\label{fig:set-up}
\end{figure*}
Our experimental setup is schematically shown in Fig.\ref{fig:set-up}. Light from a continuous wave butterfly laser diode at a wavelength of $1543.656\,$nm is amplitude modulated by using an electro-optic modulator (AM) locked to the minimum of transmission. A programmable pulse generator applies rectangular pulses carving the early (\emph{e}) and late (\emph{l}) TB modes. The delay and the pulse duration are set to $\tau = 13\,$ns and $t_w=2\,$ns, as a trade-off between a high generation rate and mitigation of free-carrier (FC) absorption induced by two-photon absorption ($\tau>\tau_{FC}$ and $t_w\ll \tau_{FC}$, where the measured FC lifetime is $\tau_{FC}\sim 10\,$ns). The repetition rate of the experiment is $50\,$MHz.\\
\indent The two pulses are then sent to a phase modulator (PM) driven by an RF signal at a frequency $f=18.25\,$GHz to produce several and coherent sidebands which are equally spaced in frequency by the modulation frequency. The time-bandwidth product is $f\cdot t_w \simeq 36 \gg 1$, which allows one to treat the TB and the FB as two \textit{distinct} DoFs, i.e. the extent of the different frequency-bin wavepackets do not overlap. After the PM, the pump light is amplified by an Erbium-Doped Fiber Amplifier (EDFA), and the background noise together with residual electro-optic comb lines are removed using a band-pass filter (BPF). A fiber polarization controller is used to set the light polarization to TE before coupling it into the photonic chip via a lensed fiber and an on-chip inverse taper (coupling loss of $\sim3.5\,$dB/facet).\\
\indent A sketch of the device is shown in Fig.\ref{fig:set-up}(a.1) (details on the fabrication procedure can be found in Ref.\cite{2023_Clementi, 2023_Borghi, 2024_Tagliavacche}). A bus waveguide ($220$ x $600\,$nm$^2$ of cross-section) couples light to a series of two nominally identical ring resonators (R1 and R2, with a radius of $22\, \mu$m and corresponding free spectral range of $\sim 524\,$ GHz). The spectra of the pump, signal and idler resonances involved in the generation of photon pairs via SFWM are reported in Fig.\ref{fig:set-up}(c). We used two thermal phase shifters, placed above the resonators, to align the pump resonance of R1 and R2 to two adjacent comb lines of the input pump (tuning efficiency $\sim0.3$ mW/deg, up to a FSR of spectral shift). The rings exhibit quality factor of $\sim10^5$ in the range of interest, thus the resulting FB spacing $f$ is approximately 100 times larger than the resonance linewidth, ensuring that the crosstalk between the frequency modes can be neglected. The intensity of the two pump lines is regulated by setting the amplitude of the driving RF signal (about $V_{pp}=9V$) to equalize the pair generation rate of the two rings, both showing an average brightness of $(1.5\pm0.1)\,\cdot10^7\,$Hz/mW$^{2}$/GHz \cite{2024_Tagliavacche}. This configuration allows each of the two pump pulses to generate a frequency-bin encoded, maximally entangled Bell state $\ket{\Phi^{+}_{FB}} = \nicefrac{1}{\sqrt{2}} \tonde{\ket{0}_s\ket{0}_i + \ket{1}_s\ket{1}_i}$ at the output of the chip, where $\ket{0}_{s(i)}$ and $\ket{1}_{s(i)}$ denote the photon state associated with the signal (idler) FB generated by respectively R1 and R2.\\
\indent Since the two identical pump pulses have equal probability of triggering the generation of photon pairs, the resulting state is the HE state: 
$\ket{\Phi^{+}_{HE}} = \ket{\Phi^{+}_{TB}} \otimes \ket{\Phi^{+}_{FB}} = \nicefrac{1}{2} \tonde{\ket{ee} + \ket{ll}}_{TB} \otimes \tonde{\ket{00} + \ket{11}}_{FB}$
where $\ket{\Phi^{+}_{TB}} = \nicefrac{1}{\sqrt{2}} \tonde{\ket{e}_s\ket{e}_i + \ket{l}_s\ket{l}_i}$. The Hilbert space of each photon is \textit{4}-dimensional, yielding a two-photon combined space of dimension \textit{16} (Fig.\ref{fig:set-up}(b)).\\
\indent The signal-idler photon pairs are coupled out through an inverse taper and collected with a lensed fiber, thus finally sent either to the time-bin analyzer (\textit{T-a}) or the frequency-bin analyzer (\textit{F-a}) reported, respectivelly, in Fig.\ref{fig:set-up}(a.2) and as Fig.\ref{fig:set-up}(a.3).\\
\indent The time-bin analyzer consists of an unbalanced fiber-based Mach-Zenhder interferometer, made by two 50:50 fiber-beam splitters and an excess fiber length between the \emph{long} and the \emph{short} path which corresponds to a delay $\tau$. The phase difference between the two arms of the interferometer is stabilized by monitoring with a photodiode (PD) a control beam in one of its output ports. This beam is derived from the continuous wave pump (before amplitude modulation, see Fig.\ref{fig:set-up}(a.1)), and is counter-propagating with respect to the signal-idler photons inside the interferometer. Phase locking is achieved by a Field Programmable Gate Array (FPGA): we implement an active feedback loop to keep constant the voltage provided by the PD by acting on the pump laser current. Since the interferometer is locked to a fixed value of the phase of the pump $\theta^p$ and not to an absolute reference, the sum of the signal and idler phases $\theta^{s(i)}$ is stabilized over time by the phase matching condition of SFWM \cite{2025_Borghi}. Finally, in the long arm of the \textit{T-a} we inserted a waveshaper (WS, Coherent WS-16000A) to separately manipulate the amplitude and phase at the signal and idler frequencies. After signal and idler separation into two distinct paths by means of a DWDM filter and pump attenuation (BPF), the photons are sent to a pair of Superconducting Nanowire Single-Photon Detectors (SNSPDs). Their arrival time are recorded relative to the trigger of the pulse generator driving the AM in Fig.\ref{fig:set-up}(a.1).\\ 
\indent To perform the quantum state analysis in the frequency domain, the idler and signal photons are sent to two distinct PMs (after their spectral separation using the DWDM and pump-laser removal through the BPF in Fig.\ref{fig:set-up}(a.3)). Each PM is driven by an RF signal at frequency $f$ coherent with the RF signal driving the PM of the pump beam, and followed by a frequency-resolved detection. The latter is implemented by two tunable fiber Bragg gratings with a pass-band of $12.5\,$GHz. The phase of the two RF signals can be independently controlled, which allows one to perform projective measurements on a superposition basis of the two frequency-bins (more details are provided in section \ref{par:Results} and in Ref. \cite{2023_Borghi}). Finally, photons are sent to the SNSPDs.

\section{Results}
\label{par:Results}

\begin{figure*}[ht]
\centering
\includegraphics[width = \textwidth]{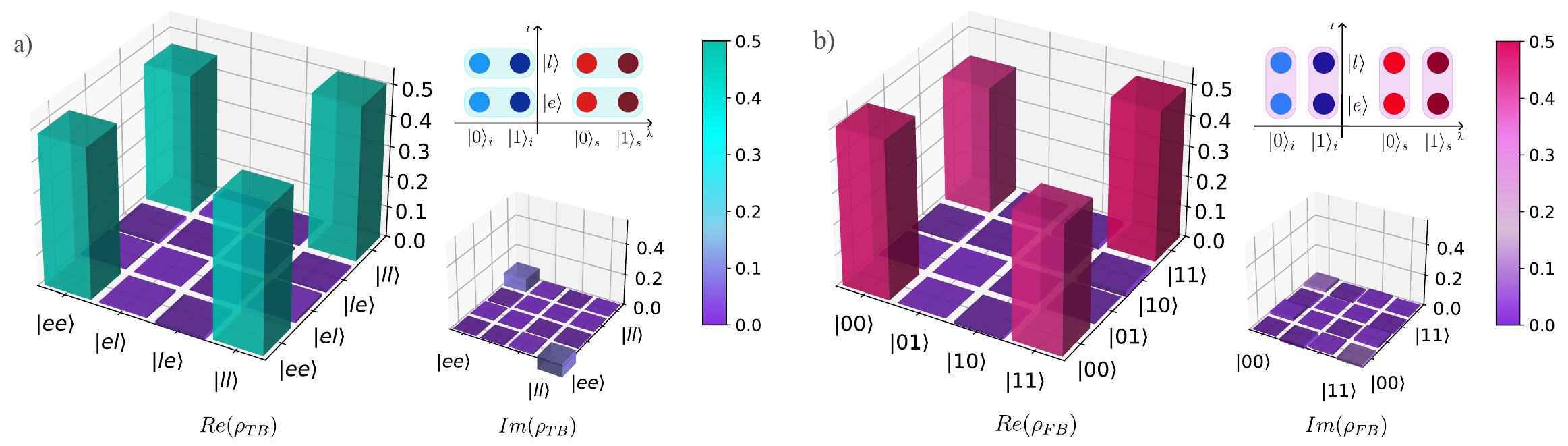}
\caption{\textbf{(a)} Reconstructed reduced density matrix (real and imaginary part) of the HE state in the TB-subspace, while tracing out the FB DoF. \textbf{(b)} Same as \textit{(a)} but referred to the FB-subspace, while tracing out the TB DoF.}
\label{fig:tomo}
\end{figure*}

To demonstrate the generation of the 16-dimensional target state $\ket{\Phi^{+}_{HE}}$, we have first proven entanglement in the TB and FB DoFs separately, and then measured an HE witness. The last step is necessary to rule out the possibility that the target state, although exhibiting entanglement in each DoF, could be written as a mixture of two or more non-HE states \cite{2008_Vallone_witness}. For all the measurements, we set the input average power to $-5.5\,$dBm, corresponding to a peak power in the bus waveguide of $350\, \mu$W for both pumps. In this regime, multiple-pair emission is negligible, and the measured coincidence to accidental ratio (CAR) is $\sim 30$. The rate of generated HE photons at the output of the chip has been measured to be $4.7\pm0.1\,$kHz.\\
\indent We performed full quantum state tomography in the time-bin and in the frequency-bin domain using, respectively, the \textit{T-a} and the \textit{F-a} setups. A complete set of projectors is formed from the following sets: $Z=\graffe{\ket{0},\ket{1}}$, $X=\graffe{\frac{1}{\sqrt{2}}(\ket{0}+\ket{1}),\frac{1}{\sqrt{2}}(\ket{0}-\ket{1})}$ and $Y=\graffe{\frac{1}{\sqrt{2}}(\ket{0}+i\ket{1}),\frac{1}{\sqrt{2}}(\ket{0}-i\ket{1})}$, where $(\ket{0},\ket{1})$ labels the logical state of either the TB or FB on each DoF \cite{takesue2009implementation}.\\
\indent The TB projectors are implemented through the \textit{T}-analyzer by post-selecting their arrival time $\Delta t_{s(i)}$ with respect to the AM trigger. For example, the detection of both photons at the delay $\Delta t_{s(i)}=0$ corresponds to a pair generated in the \emph{early} pulse and traveling the short arm of the interferometer, thus corresponding to the TB projector $\ket{ee}\bra{ee}$. Similarly, pairs generated in the \emph{late} pulse and traveling the \emph{long} path produce a coincidence at $\Delta t_{s(i)}= 2\tau$, corresponding to the TB projector $\ket{ll}\bra{ll}$. The projectors in the superposition basis $X$ and $Y$, correspond to signal-idler coincidences with both photons arriving at the detector at a time $\Delta t_{s(i)}=\tau$, in which quantum interference occurs. To span across the different superposition-basis projectors, we independently adjust the phase at the signal and idler frequencies by using the WS in the long arm, thereby setting the required phase difference between the two time bins ($Z \rightarrow 0$, $X \rightarrow \pm \pi$, $Y \rightarrow \pm \nicefrac{\pi}{2}$). At this stage, we do not resolve the frequency-bin of the signal and the idler photon, i.e. we trace out the FB DoF.\\
\indent The FB projectors are implemented by frequency-resolved detection. In particular, those in the $X_{s(i)}$ and $Y_{s(i)}$ bases involve coherent mixing the $\ket{0}_{s(i)}$ and the $\ket{1}_{s(i)}$ frequency-bin \cite{2023_Borghi}. This is achieved by activating the RF signal that drives the signal (idler) PM and equalizing the intensity of the sidebands involved in the mixing by adjusting the RF power. The trace over the TB DoF is performed, for each combination of the signal-idler frequency-bins, by summing all the events in which both the signal and the idler photons arrived at $\Delta t_{s(i)}=0$ (pairs generated in the \emph{e}-pulse), or $\Delta t_{s(i)}=\tau$ (pairs generated by \emph{l}-pulse), or in which the two photons arrived with a relative delay of $\tau$.
%%%%%%%%%%%%%%%%%% table for 1 DoF figure of merits 
\begin{table}[htbp]
\centering
\caption{\bf %List of the values of fidelity, purity and \emph{S}-parameter extracted from the reduced density matrix in each DoF
Figures of merits of the reduced density matrices}
\begin{tabularx}{\columnwidth}{c|>{\centering\arraybackslash}X>{\centering\arraybackslash}X>{\centering\arraybackslash}X}
\hline
\hline
\textbf{DoF} & Fidelity ($\%$) & Purity ($\%$) & \emph{S}-parameter \\
\hline
\textbf{FB} & $90.6 \pm 1.1$ & $83.7 \pm 1.3$ & $2.55 \pm 0.02 $ \\
\textbf{TB} & $94.9 \pm 0.6$ & $92.1 \pm 0.6$ & $2.669 \pm 0.013$ \\
\hline
\hline
\end{tabularx}
\label{tab:tomo}
\end{table}\\
\indent The reconstructed reduced density matrices are reported in Fig.\ref{fig:tomo}. More details on the reconstruction algorithm can be found in Ref.\cite{2023_Borghi}. The purity of the marginal state over the FB DoF is $(83.7\pm1.3)\%$, and is limited by the imperfect spectral indistinguishability of the photon pairs generated by the two rings \cite{2023_Borghi}. We support this evidence by independently measuring the visibility of two-photon interference in both the two DoF subspaces, occurring when the signal and the idler photons are jointly measured in a superposition basis, namely $\nicefrac{1}{\sqrt{2}}(\ket{0}+e^{i\phi}\ket{1})$, and the phase $\phi$ is swept.
% the signal and the idler photons are projected along the equator of their local Bloch sphere and the azimuthal angle is swept. 
We found a visibility of $(83.4\pm 1.4)\%$, which also coincides with the indistinguishability between the two photon pair sources. The purity of the state in the TB subspace is $(92.1\pm0.6)\%$, and is consistent with the measured two photon interference visibility of $(93\pm 1)\%$. The latter is independently measured by sweeping the phase of the signal photon in the \textit{T-a}, and by post-selecting the coincidence events where both photons arrived at the detectors at a relative time $\Delta t_{s(i)}=\tau$ with respect to the pump trigger. The non-unit purity in the TB subspace is likely attributed to thermal and free-carrier relaxation occurring between the two consecutive pump pulses. The resonance wavelength and the amount of free-carrier absorption slightly differs from the early and the late pulses, thereby causing distinguishability in the joint temporal emission of photon pairs between consecutive pulses. The FB and TB marginal states were prepared with fidelity exceeding $90 \%$ with respect to the maximally entangled Bell states $\ket{\Phi^{+}_{FB}}$ and $\ket{\Phi^{+}_{TB}}$ respectively (see Table \ref{tab:tomo}). From the reduced density matrices we calculated the value of the CHSH inequality (\emph{S}-parameter), proving the presence of entanglement by exceeding the maximum value of $S=2$ for separable states by more than 27 STDs in both the subspaces. These values are limited by the non-unit purity of the marginal states.\\
\indent As a final step, we verified the presence of HE in the TB and FB DoF by measuring the witness \cite{2008_Vallone_witness}: $W = (N-1)-\sum^N_{k=1} S_k\,$,
where $N=2\cdot d$, $d$ is the number of DoFs and $S_k$ are the \emph{stabilizers} of the HE state, which in our specific case are
$S_{1} = \sigma_{X \, s}^{TB} \, \sigma_{X \, i}^{TB}=0.905\pm0.005 \, , 
\, S_{2} = \sigma_{Z \, s}^{TB} \, \sigma_{Z \, i}^{TB} =0.981\pm0.004\, , 
\, S_{3} = \sigma_{X \, s}^{FB} \, \sigma_{X \, i}^{FB} =0.720 \pm 0.007\, , 
\,  S_{4} = \sigma_{Z \, s}^{FB} \, \sigma_{Z \, i}^{FB} =0.994 \pm 0.001$. The Pauli matrices $\sigma_{X_j,Z_j}^{TB(FB)}$ acts on the $j=\{s,i\}$ photon of the TB (FB) subspace. The defined witness $W$ confirms the presence of HE whenever $\langle W \rangle < 0$. Experimentally, we measured a value of $-0.60\pm0.01$, being lower than the minimum value for non-HE states ($\langle W\rangle \ge0$) by 60 STDs. 

\section{Conclusions}
We have demonstrated the generation of hyperentangled photon pairs from a silicon-on-insulator source using the time-bin and the frequency-bin encoding. The present geometry, consisting of two high-brightness coherently driven microring resonators, allows one to achieve both high photon-pairs generation rate and small frequency-bin spacing, overcoming a fundamental trade-off inherent to single-resonator systems.
To characterize the generated state, we implemented time-bin and frequency-bin analyzers using standard fiber optic components, and reconstructed the reduced density matrices for both degrees of freedom. Through CHSH inequality violations, we proved the effective generation of entangled states in each DoF, and further employed a HE witness to confirm the presence of genuine hyperentanglement, ruling out the possibility of a statistical mixture of non-hyperentangled states.\\
\indent Higher-dimensional entangled states can be generated with our approach with minimal additional resources, for instance, by increasing the number of microring resonators \cite{2023_Borghi} or the number of pump pulses \cite{2025_Morendotti_TB8}. In next generations of the device, we envision incorporating other auxiliary DoFs, such as path or polarization. 
The ability to encode quantum information in time-bin and frequency-bin DoFs is particularly well suited for quantum communication applications over free-space and fiber-based channels. Moreover, the compatibility of our approach with silicon-on-insulator technology and standard telecom-grade components highlights its potential for scalable and practical quantum communication systems.

\section*{FUNDING AND ACKNOWLEDGEMENTS}
D.B. acknowledges the support of the Italian MUR and the European Union—Next Generation EU through the PRIN project number F53D23000550006—SIGNED. M.B., E.B, and M.L. acknowledge the PNRR MUR project PE0000023-NQSTI. S.O. and M.G. acknowledge European Union funding from the HyperSpace project (project ID 101070168).
\\This work was partly supported by the French RENATECH network. 

% Bibliography
%%\bibliography{reference}

\begin{thebibliography}{37}%
\makeatletter
\providecommand \@ifxundefined [1]{%
 \@ifx{#1\undefined}
}%
\providecommand \@ifnum [1]{%
 \ifnum #1\expandafter \@firstoftwo
 \else \expandafter \@secondoftwo
 \fi
}%
\providecommand \@ifx [1]{%
 \ifx #1\expandafter \@firstoftwo
 \else \expandafter \@secondoftwo
 \fi
}%
\providecommand \natexlab [1]{#1}%
\providecommand \enquote  [1]{``#1''}%
\providecommand \bibnamefont  [1]{#1}%
\providecommand \bibfnamefont [1]{#1}%
\providecommand \citenamefont [1]{#1}%
\providecommand \href@noop [0]{\@secondoftwo}%
\providecommand \href [0]{\begingroup \@sanitize@url \@href}%
\providecommand \@href[1]{\@@startlink{#1}\@@href}%
\providecommand \@@href[1]{\endgroup#1\@@endlink}%
\providecommand \@sanitize@url [0]{\catcode `\\12\catcode `\$12\catcode `\&12\catcode `\#12\catcode `\^12\catcode `\_12\catcode `\%12\relax}%
\providecommand \@@startlink[1]{}%
\providecommand \@@endlink[0]{}%
\providecommand \url  [0]{\begingroup\@sanitize@url \@url }%
\providecommand \@url [1]{\endgroup\@href {#1}{\urlprefix }}%
\providecommand \urlprefix  [0]{URL }%
\providecommand \Eprint [0]{\href }%
\providecommand \doibase [0]{https://doi.org/}%
\providecommand \selectlanguage [0]{\@gobble}%
\providecommand \bibinfo  [0]{\@secondoftwo}%
\providecommand \bibfield  [0]{\@secondoftwo}%
\providecommand \translation [1]{[#1]}%
\providecommand \BibitemOpen [0]{}%
\providecommand \bibitemStop [0]{}%
\providecommand \bibitemNoStop [0]{.\EOS\space}%
\providecommand \EOS [0]{\spacefactor3000\relax}%
\providecommand \BibitemShut  [1]{\csname bibitem#1\endcsname}%
\let\auto@bib@innerbib\@empty
%</preamble>
\bibitem [{\citenamefont {Kwiat}(1997)}]{1997_Kwiat_HE}%
  \BibitemOpen
  \bibfield  {author} {\bibinfo {author} {\bibfnamefont {P.~G.}\ \bibnamefont {Kwiat}},\ }\bibfield  {title} {\bibinfo {title} {Hyper-entangled states},\ }\href@noop {} {\bibfield  {journal} {\bibinfo  {journal} {Journal of modern optics}\ }\textbf {\bibinfo {volume} {44}},\ \bibinfo {pages} {2173} (\bibinfo {year} {1997})}\BibitemShut {NoStop}%
\bibitem [{\citenamefont {Schuck}\ \emph {et~al.}(2006)\citenamefont {Schuck}, \citenamefont {Huber}, \citenamefont {Kurtsiefer},\ and\ \citenamefont {Weinfurter}}]{2006_Schuck_HBSA}%
  \BibitemOpen
  \bibfield  {author} {\bibinfo {author} {\bibfnamefont {C.}~\bibnamefont {Schuck}}, \bibinfo {author} {\bibfnamefont {G.}~\bibnamefont {Huber}}, \bibinfo {author} {\bibfnamefont {C.}~\bibnamefont {Kurtsiefer}},\ and\ \bibinfo {author} {\bibfnamefont {H.}~\bibnamefont {Weinfurter}},\ }\bibfield  {title} {\bibinfo {title} {Complete deterministic linear optics bell state analysis},\ }\href@noop {} {\bibfield  {journal} {\bibinfo  {journal} {Physical review letters}\ }\textbf {\bibinfo {volume} {96}},\ \bibinfo {pages} {190501} (\bibinfo {year} {2006})}\BibitemShut {NoStop}%
\bibitem [{\citenamefont {Wei}\ \emph {et~al.}(2007)\citenamefont {Wei}, \citenamefont {Barreiro},\ and\ \citenamefont {Kwiat}}]{2007_Wei_HBSA}%
  \BibitemOpen
  \bibfield  {author} {\bibinfo {author} {\bibfnamefont {T.-C.}\ \bibnamefont {Wei}}, \bibinfo {author} {\bibfnamefont {J.~T.}\ \bibnamefont {Barreiro}},\ and\ \bibinfo {author} {\bibfnamefont {P.~G.}\ \bibnamefont {Kwiat}},\ }\bibfield  {title} {\bibinfo {title} {Hyperentangled bell-state analysis},\ }\href@noop {} {\bibfield  {journal} {\bibinfo  {journal} {Physical Review A-Atomic, Molecular, and Optical Physics}\ }\textbf {\bibinfo {volume} {75}},\ \bibinfo {pages} {060305} (\bibinfo {year} {2007})}\BibitemShut {NoStop}%
\bibitem [{\citenamefont {Barreiro}\ \emph {et~al.}(2008)\citenamefont {Barreiro}, \citenamefont {Wei},\ and\ \citenamefont {Kwiat}}]{2008_Barreiro_superdence}%
  \BibitemOpen
  \bibfield  {author} {\bibinfo {author} {\bibfnamefont {J.~T.}\ \bibnamefont {Barreiro}}, \bibinfo {author} {\bibfnamefont {T.-C.}\ \bibnamefont {Wei}},\ and\ \bibinfo {author} {\bibfnamefont {P.~G.}\ \bibnamefont {Kwiat}},\ }\bibfield  {title} {\bibinfo {title} {Beating the channel capacity limit for linear photonic superdense coding},\ }\href@noop {} {\bibfield  {journal} {\bibinfo  {journal} {Nature physics}\ }\textbf {\bibinfo {volume} {4}},\ \bibinfo {pages} {282} (\bibinfo {year} {2008})}\BibitemShut {NoStop}%
\bibitem [{\citenamefont {Williams}\ \emph {et~al.}(2017)\citenamefont {Williams}, \citenamefont {Sadlier},\ and\ \citenamefont {Humble}}]{2017_williams_superdense}%
  \BibitemOpen
  \bibfield  {author} {\bibinfo {author} {\bibfnamefont {B.~P.}\ \bibnamefont {Williams}}, \bibinfo {author} {\bibfnamefont {R.~J.}\ \bibnamefont {Sadlier}},\ and\ \bibinfo {author} {\bibfnamefont {T.~S.}\ \bibnamefont {Humble}},\ }\bibfield  {title} {\bibinfo {title} {Superdense coding over optical fiber links with complete bell-state measurements},\ }\href@noop {} {\bibfield  {journal} {\bibinfo  {journal} {Physical review letters}\ }\textbf {\bibinfo {volume} {118}},\ \bibinfo {pages} {050501} (\bibinfo {year} {2017})}\BibitemShut {NoStop}%
\bibitem [{\citenamefont {Chen}\ \emph {et~al.}(2007)\citenamefont {Chen}, \citenamefont {Li}, \citenamefont {Zhang}, \citenamefont {Chen}, \citenamefont {Goebel}, \citenamefont {Chen}, \citenamefont {Mair},\ and\ \citenamefont {Pan}}]{2007_Chen_HEtoCluster}%
  \BibitemOpen
  \bibfield  {author} {\bibinfo {author} {\bibfnamefont {K.}~\bibnamefont {Chen}}, \bibinfo {author} {\bibfnamefont {C.-M.}\ \bibnamefont {Li}}, \bibinfo {author} {\bibfnamefont {Q.}~\bibnamefont {Zhang}}, \bibinfo {author} {\bibfnamefont {Y.-A.}\ \bibnamefont {Chen}}, \bibinfo {author} {\bibfnamefont {A.}~\bibnamefont {Goebel}}, \bibinfo {author} {\bibfnamefont {S.}~\bibnamefont {Chen}}, \bibinfo {author} {\bibfnamefont {A.}~\bibnamefont {Mair}},\ and\ \bibinfo {author} {\bibfnamefont {J.-W.}\ \bibnamefont {Pan}},\ }\bibfield  {title} {\bibinfo {title} {Experimental realization of one-way quantum computing with two-photon four-qubit cluster states},\ }\href@noop {} {\bibfield  {journal} {\bibinfo  {journal} {Physical review letters}\ }\textbf {\bibinfo {volume} {99}},\ \bibinfo {pages} {120503} (\bibinfo {year} {2007})}\BibitemShut {NoStop}%
\bibitem [{\citenamefont {Vallone}\ \emph {et~al.}(2008{\natexlab{a}})\citenamefont {Vallone}, \citenamefont {Pomarico}, \citenamefont {De~Martini},\ and\ \citenamefont {Mataloni}}]{2008_Vallone_HEtoCluster}%
  \BibitemOpen
  \bibfield  {author} {\bibinfo {author} {\bibfnamefont {G.}~\bibnamefont {Vallone}}, \bibinfo {author} {\bibfnamefont {E.}~\bibnamefont {Pomarico}}, \bibinfo {author} {\bibfnamefont {F.}~\bibnamefont {De~Martini}},\ and\ \bibinfo {author} {\bibfnamefont {P.}~\bibnamefont {Mataloni}},\ }\bibfield  {title} {\bibinfo {title} {One-way quantum computation with two-photon multiqubit cluster states},\ }\href@noop {} {\bibfield  {journal} {\bibinfo  {journal} {Physical Review A-Atomic, Molecular, and Optical Physics}\ }\textbf {\bibinfo {volume} {78}},\ \bibinfo {pages} {042335} (\bibinfo {year} {2008}{\natexlab{a}})}\BibitemShut {NoStop}%
\bibitem [{\citenamefont {Reimer}\ \emph {et~al.}(2019)\citenamefont {Reimer}, \citenamefont {Sciara}, \citenamefont {Roztocki}, \citenamefont {Islam}, \citenamefont {Romero~Cort{\'e}s}, \citenamefont {Zhang}, \citenamefont {Fischer}, \citenamefont {Loranger}, \citenamefont {Kashyap}, \citenamefont {Cino} \emph {et~al.}}]{2019_Morandotti_HEtimefreq}%
  \BibitemOpen
  \bibfield  {author} {\bibinfo {author} {\bibfnamefont {C.}~\bibnamefont {Reimer}}, \bibinfo {author} {\bibfnamefont {S.}~\bibnamefont {Sciara}}, \bibinfo {author} {\bibfnamefont {P.}~\bibnamefont {Roztocki}}, \bibinfo {author} {\bibfnamefont {M.}~\bibnamefont {Islam}}, \bibinfo {author} {\bibfnamefont {L.}~\bibnamefont {Romero~Cort{\'e}s}}, \bibinfo {author} {\bibfnamefont {Y.}~\bibnamefont {Zhang}}, \bibinfo {author} {\bibfnamefont {B.}~\bibnamefont {Fischer}}, \bibinfo {author} {\bibfnamefont {S.}~\bibnamefont {Loranger}}, \bibinfo {author} {\bibfnamefont {R.}~\bibnamefont {Kashyap}}, \bibinfo {author} {\bibfnamefont {A.}~\bibnamefont {Cino}}, \emph {et~al.},\ }\bibfield  {title} {\bibinfo {title} {High-dimensional one-way quantum processing implemented on d-level cluster states},\ }\href@noop {} {\bibfield  {journal} {\bibinfo  {journal} {Nature Physics}\ }\textbf {\bibinfo {volume} {15}},\ \bibinfo {pages} {148} (\bibinfo {year} {2019})}\BibitemShut {NoStop}%
\bibitem [{\citenamefont {Kim}\ \emph {et~al.}(2021)\citenamefont {Kim}, \citenamefont {Kim}, \citenamefont {Im}, \citenamefont {Lee}, \citenamefont {Chae}, \citenamefont {Scarcelli},\ and\ \citenamefont {Kim}}]{2021_Kim_noiseresistant}%
  \BibitemOpen
  \bibfield  {author} {\bibinfo {author} {\bibfnamefont {J.-H.}\ \bibnamefont {Kim}}, \bibinfo {author} {\bibfnamefont {Y.}~\bibnamefont {Kim}}, \bibinfo {author} {\bibfnamefont {D.-G.}\ \bibnamefont {Im}}, \bibinfo {author} {\bibfnamefont {C.-H.}\ \bibnamefont {Lee}}, \bibinfo {author} {\bibfnamefont {J.-W.}\ \bibnamefont {Chae}}, \bibinfo {author} {\bibfnamefont {G.}~\bibnamefont {Scarcelli}},\ and\ \bibinfo {author} {\bibfnamefont {Y.-H.}\ \bibnamefont {Kim}},\ }\bibfield  {title} {\bibinfo {title} {Noise-resistant quantum communications using hyperentanglement},\ }\href@noop {} {\bibfield  {journal} {\bibinfo  {journal} {Optica}\ }\textbf {\bibinfo {volume} {8}},\ \bibinfo {pages} {1524} (\bibinfo {year} {2021})}\BibitemShut {NoStop}%
\bibitem [{\citenamefont {Xu}\ \emph {et~al.}(2023)\citenamefont {Xu}, \citenamefont {Chen}, \citenamefont {Kirby},\ and\ \citenamefont {Qian}}]{2023_Xu_Distillation_PolFreq}%
  \BibitemOpen
  \bibfield  {author} {\bibinfo {author} {\bibfnamefont {D.}~\bibnamefont {Xu}}, \bibinfo {author} {\bibfnamefont {C.}~\bibnamefont {Chen}}, \bibinfo {author} {\bibfnamefont {B.~T.}\ \bibnamefont {Kirby}},\ and\ \bibinfo {author} {\bibfnamefont {L.}~\bibnamefont {Qian}},\ }\bibfield  {title} {\bibinfo {title} {Entanglement distillation based on polarization and frequency hyperentanglement},\ }\href@noop {} {\bibfield  {journal} {\bibinfo  {journal} {IEEE Journal of Selected Topics in Quantum Electronics}\ }\textbf {\bibinfo {volume} {30}},\ \bibinfo {pages} {1} (\bibinfo {year} {2023})}\BibitemShut {NoStop}%
\bibitem [{\citenamefont {Deng}\ \emph {et~al.}(2017)\citenamefont {Deng}, \citenamefont {Ren},\ and\ \citenamefont {Li}}]{2017_Deng_HE}%
  \BibitemOpen
  \bibfield  {author} {\bibinfo {author} {\bibfnamefont {F.-G.}\ \bibnamefont {Deng}}, \bibinfo {author} {\bibfnamefont {B.-C.}\ \bibnamefont {Ren}},\ and\ \bibinfo {author} {\bibfnamefont {X.-H.}\ \bibnamefont {Li}},\ }\bibfield  {title} {\bibinfo {title} {Quantum hyperentanglement and its applications in quantum information processing},\ }\href@noop {} {\bibfield  {journal} {\bibinfo  {journal} {Science bulletin}\ }\textbf {\bibinfo {volume} {62}},\ \bibinfo {pages} {46} (\bibinfo {year} {2017})}\BibitemShut {NoStop}%
\bibitem [{\citenamefont {Barreiro}\ \emph {et~al.}(2005)\citenamefont {Barreiro}, \citenamefont {Langford}, \citenamefont {Peters},\ and\ \citenamefont {Kwiat}}]{2005_Barreiro_3DoFs}%
  \BibitemOpen
  \bibfield  {author} {\bibinfo {author} {\bibfnamefont {J.~T.}\ \bibnamefont {Barreiro}}, \bibinfo {author} {\bibfnamefont {N.~K.}\ \bibnamefont {Langford}}, \bibinfo {author} {\bibfnamefont {N.~A.}\ \bibnamefont {Peters}},\ and\ \bibinfo {author} {\bibfnamefont {P.~G.}\ \bibnamefont {Kwiat}},\ }\bibfield  {title} {\bibinfo {title} {Generation of hyperentangled photon pairs},\ }\href {https://doi.org/10.1103/PhysRevLett.95.260501} {\bibfield  {journal} {\bibinfo  {journal} {Phys. Rev. Lett.}\ }\textbf {\bibinfo {volume} {95}},\ \bibinfo {pages} {260501} (\bibinfo {year} {2005})}\BibitemShut {NoStop}%
\bibitem [{\citenamefont {Ceccarelli}\ \emph {et~al.}(2009)\citenamefont {Ceccarelli}, \citenamefont {Vallone}, \citenamefont {De~Martini}, \citenamefont {Mataloni},\ and\ \citenamefont {Cabello}}]{2009_Ceccarelli_Vallone_3DoFs}%
  \BibitemOpen
  \bibfield  {author} {\bibinfo {author} {\bibfnamefont {R.}~\bibnamefont {Ceccarelli}}, \bibinfo {author} {\bibfnamefont {G.}~\bibnamefont {Vallone}}, \bibinfo {author} {\bibfnamefont {F.}~\bibnamefont {De~Martini}}, \bibinfo {author} {\bibfnamefont {P.}~\bibnamefont {Mataloni}},\ and\ \bibinfo {author} {\bibfnamefont {A.}~\bibnamefont {Cabello}},\ }\bibfield  {title} {\bibinfo {title} {Experimental entanglement and nonlocality of a two-photon six-qubit cluster state},\ }\href {https://doi.org/10.1103/PhysRevLett.103.160401} {\bibfield  {journal} {\bibinfo  {journal} {Phys. Rev. Lett.}\ }\textbf {\bibinfo {volume} {103}},\ \bibinfo {pages} {160401} (\bibinfo {year} {2009})}\BibitemShut {NoStop}%
\bibitem [{\citenamefont {Vallone}\ \emph {et~al.}(2009)\citenamefont {Vallone}, \citenamefont {Ceccarelli}, \citenamefont {De~Martini},\ and\ \citenamefont {Mataloni}}]{2009_Vallone_Ceccarelli_3DoFs}%
  \BibitemOpen
  \bibfield  {author} {\bibinfo {author} {\bibfnamefont {G.}~\bibnamefont {Vallone}}, \bibinfo {author} {\bibfnamefont {R.}~\bibnamefont {Ceccarelli}}, \bibinfo {author} {\bibfnamefont {F.}~\bibnamefont {De~Martini}},\ and\ \bibinfo {author} {\bibfnamefont {P.}~\bibnamefont {Mataloni}},\ }\bibfield  {title} {\bibinfo {title} {Hyperentanglement of two photons in three degrees of freedom},\ }\href@noop {} {\bibfield  {journal} {\bibinfo  {journal} {Physical Review A-Atomic, Molecular, and Optical Physics}\ }\textbf {\bibinfo {volume} {79}},\ \bibinfo {pages} {030301} (\bibinfo {year} {2009})}\BibitemShut {NoStop}%
\bibitem [{\citenamefont {Lu}\ \emph {et~al.}(2023{\natexlab{a}})\citenamefont {Lu}, \citenamefont {Alshowkan}, \citenamefont {Myilswamy}, \citenamefont {Weiner}, \citenamefont {Lukens},\ and\ \citenamefont {Peters}}]{2023_Lu_ultrabroad}%
  \BibitemOpen
  \bibfield  {author} {\bibinfo {author} {\bibfnamefont {H.-H.}\ \bibnamefont {Lu}}, \bibinfo {author} {\bibfnamefont {M.}~\bibnamefont {Alshowkan}}, \bibinfo {author} {\bibfnamefont {K.~V.}\ \bibnamefont {Myilswamy}}, \bibinfo {author} {\bibfnamefont {A.~M.}\ \bibnamefont {Weiner}}, \bibinfo {author} {\bibfnamefont {J.~M.}\ \bibnamefont {Lukens}},\ and\ \bibinfo {author} {\bibfnamefont {N.~A.}\ \bibnamefont {Peters}},\ }\bibfield  {title} {\bibinfo {title} {Generation and characterization of ultrabroadband polarization--frequency hyperentangled photons},\ }\href@noop {} {\bibfield  {journal} {\bibinfo  {journal} {Optics Letters}\ }\textbf {\bibinfo {volume} {48}},\ \bibinfo {pages} {6031} (\bibinfo {year} {2023}{\natexlab{a}})}\BibitemShut {NoStop}%
\bibitem [{\citenamefont {Xie}\ \emph {et~al.}(2015)\citenamefont {Xie}, \citenamefont {Zhong}, \citenamefont {Shrestha}, \citenamefont {Xu}, \citenamefont {Liang}, \citenamefont {Gong}, \citenamefont {Bienfang}, \citenamefont {Restelli}, \citenamefont {Shapiro}, \citenamefont {Wong} \emph {et~al.}}]{2015_Xie_5qubit}%
  \BibitemOpen
  \bibfield  {author} {\bibinfo {author} {\bibfnamefont {Z.}~\bibnamefont {Xie}}, \bibinfo {author} {\bibfnamefont {T.}~\bibnamefont {Zhong}}, \bibinfo {author} {\bibfnamefont {S.}~\bibnamefont {Shrestha}}, \bibinfo {author} {\bibfnamefont {X.}~\bibnamefont {Xu}}, \bibinfo {author} {\bibfnamefont {J.}~\bibnamefont {Liang}}, \bibinfo {author} {\bibfnamefont {Y.-X.}\ \bibnamefont {Gong}}, \bibinfo {author} {\bibfnamefont {J.~C.}\ \bibnamefont {Bienfang}}, \bibinfo {author} {\bibfnamefont {A.}~\bibnamefont {Restelli}}, \bibinfo {author} {\bibfnamefont {J.~H.}\ \bibnamefont {Shapiro}}, \bibinfo {author} {\bibfnamefont {F.~N.}\ \bibnamefont {Wong}}, \emph {et~al.},\ }\bibfield  {title} {\bibinfo {title} {Harnessing high-dimensional hyperentanglement through a biphoton frequency comb},\ }\href@noop {} {\bibfield  {journal} {\bibinfo  {journal} {Nature Photonics}\ }\textbf {\bibinfo {volume} {9}},\ \bibinfo {pages} {536} (\bibinfo {year} {2015})}\BibitemShut {NoStop}%
\bibitem [{\citenamefont {Gao}\ \emph {et~al.}(2010)\citenamefont {Gao}, \citenamefont {Lu}, \citenamefont {Yao}, \citenamefont {Xu}, \citenamefont {G{\"u}hne}, \citenamefont {Goebel}, \citenamefont {Chen}, \citenamefont {Peng}, \citenamefont {Chen},\ and\ \citenamefont {Pan}}]{2010_Gao_Cat}%
  \BibitemOpen
  \bibfield  {author} {\bibinfo {author} {\bibfnamefont {W.-B.}\ \bibnamefont {Gao}}, \bibinfo {author} {\bibfnamefont {C.-Y.}\ \bibnamefont {Lu}}, \bibinfo {author} {\bibfnamefont {X.-C.}\ \bibnamefont {Yao}}, \bibinfo {author} {\bibfnamefont {P.}~\bibnamefont {Xu}}, \bibinfo {author} {\bibfnamefont {O.}~\bibnamefont {G{\"u}hne}}, \bibinfo {author} {\bibfnamefont {A.}~\bibnamefont {Goebel}}, \bibinfo {author} {\bibfnamefont {Y.-A.}\ \bibnamefont {Chen}}, \bibinfo {author} {\bibfnamefont {C.-Z.}\ \bibnamefont {Peng}}, \bibinfo {author} {\bibfnamefont {Z.-B.}\ \bibnamefont {Chen}},\ and\ \bibinfo {author} {\bibfnamefont {J.-W.}\ \bibnamefont {Pan}},\ }\bibfield  {title} {\bibinfo {title} {Experimental demonstration of a hyper-entangled ten-qubit schr{\"o}dinger cat state},\ }\href@noop {} {\bibfield  {journal} {\bibinfo  {journal} {Nature physics}\ }\textbf {\bibinfo {volume} {6}},\ \bibinfo {pages} {331} (\bibinfo {year} {2010})}\BibitemShut {NoStop}%
\bibitem [{\citenamefont {Chen}\ \emph {et~al.}(2021)\citenamefont {Chen}, \citenamefont {Fu}, \citenamefont {Gong},\ and\ \citenamefont {Wang}}]{2021_Chen_review}%
  \BibitemOpen
  \bibfield  {author} {\bibinfo {author} {\bibfnamefont {X.}~\bibnamefont {Chen}}, \bibinfo {author} {\bibfnamefont {Z.}~\bibnamefont {Fu}}, \bibinfo {author} {\bibfnamefont {Q.}~\bibnamefont {Gong}},\ and\ \bibinfo {author} {\bibfnamefont {J.}~\bibnamefont {Wang}},\ }\bibfield  {title} {\bibinfo {title} {Quantum entanglement on photonic chips: a review},\ }\href@noop {} {\bibfield  {journal} {\bibinfo  {journal} {Advanced Photonics}\ }\textbf {\bibinfo {volume} {3}},\ \bibinfo {pages} {064002} (\bibinfo {year} {2021})}\BibitemShut {NoStop}%
\bibitem [{\citenamefont {Kumar}\ \emph {et~al.}(2020)\citenamefont {Kumar}, \citenamefont {Yadav},\ and\ \citenamefont {Ghosh}}]{2020_Kumar_LiNbO3}%
  \BibitemOpen
  \bibfield  {author} {\bibinfo {author} {\bibfnamefont {R.}~\bibnamefont {Kumar}}, \bibinfo {author} {\bibfnamefont {V.~K.}\ \bibnamefont {Yadav}},\ and\ \bibinfo {author} {\bibfnamefont {J.}~\bibnamefont {Ghosh}},\ }\bibfield  {title} {\bibinfo {title} {Postselection-free, hyperentangled photon pairs in a periodically poled lithium-niobate ridge waveguide},\ }\href@noop {} {\bibfield  {journal} {\bibinfo  {journal} {Physical Review A}\ }\textbf {\bibinfo {volume} {102}},\ \bibinfo {pages} {033722} (\bibinfo {year} {2020})}\BibitemShut {NoStop}%
\bibitem [{\citenamefont {Huang}\ \emph {et~al.}(2022)\citenamefont {Huang}, \citenamefont {Feng}, \citenamefont {Li}, \citenamefont {Qi}, \citenamefont {Lu}, \citenamefont {Zheng},\ and\ \citenamefont {Chen}}]{2022_Huang_LiNbO3}%
  \BibitemOpen
  \bibfield  {author} {\bibinfo {author} {\bibfnamefont {Y.}~\bibnamefont {Huang}}, \bibinfo {author} {\bibfnamefont {J.}~\bibnamefont {Feng}}, \bibinfo {author} {\bibfnamefont {Y.}~\bibnamefont {Li}}, \bibinfo {author} {\bibfnamefont {Z.}~\bibnamefont {Qi}}, \bibinfo {author} {\bibfnamefont {C.}~\bibnamefont {Lu}}, \bibinfo {author} {\bibfnamefont {Y.}~\bibnamefont {Zheng}},\ and\ \bibinfo {author} {\bibfnamefont {X.}~\bibnamefont {Chen}},\ }\bibfield  {title} {\bibinfo {title} {High-performance hyperentanglement generation and manipulation based on lithium niobate waveguides},\ }\href@noop {} {\bibfield  {journal} {\bibinfo  {journal} {Physical Review Applied}\ }\textbf {\bibinfo {volume} {17}},\ \bibinfo {pages} {054002} (\bibinfo {year} {2022})}\BibitemShut {NoStop}%
\bibitem [{\citenamefont {Suo}\ \emph {et~al.}(2015)\citenamefont {Suo}, \citenamefont {Dong}, \citenamefont {Zhang}, \citenamefont {Huang},\ and\ \citenamefont {Peng}}]{2015_Suo_HEpolatime}%
  \BibitemOpen
  \bibfield  {author} {\bibinfo {author} {\bibfnamefont {J.}~\bibnamefont {Suo}}, \bibinfo {author} {\bibfnamefont {S.}~\bibnamefont {Dong}}, \bibinfo {author} {\bibfnamefont {W.}~\bibnamefont {Zhang}}, \bibinfo {author} {\bibfnamefont {Y.}~\bibnamefont {Huang}},\ and\ \bibinfo {author} {\bibfnamefont {J.}~\bibnamefont {Peng}},\ }\bibfield  {title} {\bibinfo {title} {Generation of hyper-entanglement on polarization and energy-time based on a silicon micro-ring cavity},\ }\href {https://doi.org/10.1364/OE.23.003985} {\bibfield  {journal} {\bibinfo  {journal} {Opt. Express}\ }\textbf {\bibinfo {volume} {23}},\ \bibinfo {pages} {3985} (\bibinfo {year} {2015})}\BibitemShut {NoStop}%
\bibitem [{\citenamefont {Marcikic}\ \emph {et~al.}(2004)\citenamefont {Marcikic}, \citenamefont {De~Riedmatten}, \citenamefont {Tittel}, \citenamefont {Zbinden}, \citenamefont {Legr{\'e}},\ and\ \citenamefont {Gisin}}]{2004_Marcikic_TB50km}%
  \BibitemOpen
  \bibfield  {author} {\bibinfo {author} {\bibfnamefont {I.}~\bibnamefont {Marcikic}}, \bibinfo {author} {\bibfnamefont {H.}~\bibnamefont {De~Riedmatten}}, \bibinfo {author} {\bibfnamefont {W.}~\bibnamefont {Tittel}}, \bibinfo {author} {\bibfnamefont {H.}~\bibnamefont {Zbinden}}, \bibinfo {author} {\bibfnamefont {M.}~\bibnamefont {Legr{\'e}}},\ and\ \bibinfo {author} {\bibfnamefont {N.}~\bibnamefont {Gisin}},\ }\bibfield  {title} {\bibinfo {title} {Distribution of time-bin entangled qubits over 50 km of optical fiber},\ }\href@noop {} {\bibfield  {journal} {\bibinfo  {journal} {Physical review letters}\ }\textbf {\bibinfo {volume} {93}},\ \bibinfo {pages} {180502} (\bibinfo {year} {2004})}\BibitemShut {NoStop}%
\bibitem [{\citenamefont {Ursin}\ \emph {et~al.}(2007)\citenamefont {Ursin}, \citenamefont {Tiefenbacher}, \citenamefont {Schmitt-Manderbach}, \citenamefont {Weier}, \citenamefont {Scheidl}, \citenamefont {Lindenthal}, \citenamefont {Blauensteiner}, \citenamefont {Jennewein}, \citenamefont {Perdigues}, \citenamefont {Trojek} \emph {et~al.}}]{2007_Ursin_144km}%
  \BibitemOpen
  \bibfield  {author} {\bibinfo {author} {\bibfnamefont {R.}~\bibnamefont {Ursin}}, \bibinfo {author} {\bibfnamefont {F.}~\bibnamefont {Tiefenbacher}}, \bibinfo {author} {\bibfnamefont {T.}~\bibnamefont {Schmitt-Manderbach}}, \bibinfo {author} {\bibfnamefont {H.}~\bibnamefont {Weier}}, \bibinfo {author} {\bibfnamefont {T.}~\bibnamefont {Scheidl}}, \bibinfo {author} {\bibfnamefont {M.}~\bibnamefont {Lindenthal}}, \bibinfo {author} {\bibfnamefont {B.}~\bibnamefont {Blauensteiner}}, \bibinfo {author} {\bibfnamefont {T.}~\bibnamefont {Jennewein}}, \bibinfo {author} {\bibfnamefont {J.}~\bibnamefont {Perdigues}}, \bibinfo {author} {\bibfnamefont {P.}~\bibnamefont {Trojek}}, \emph {et~al.},\ }\bibfield  {title} {\bibinfo {title} {Entanglement-based quantum communication over 144 km},\ }\href@noop {} {\bibfield  {journal} {\bibinfo  {journal} {Nature physics}\ }\textbf {\bibinfo {volume} {3}},\ \bibinfo {pages} {481} (\bibinfo {year} {2007})}\BibitemShut {NoStop}%
\bibitem [{\citenamefont {Vallone}\ \emph {et~al.}(2016)\citenamefont {Vallone}, \citenamefont {Dequal}, \citenamefont {Tomasin}, \citenamefont {Vedovato}, \citenamefont {Schiavon}, \citenamefont {Luceri}, \citenamefont {Bianco},\ and\ \citenamefont {Villoresi}}]{2016_Vallone_space}%
  \BibitemOpen
  \bibfield  {author} {\bibinfo {author} {\bibfnamefont {G.}~\bibnamefont {Vallone}}, \bibinfo {author} {\bibfnamefont {D.}~\bibnamefont {Dequal}}, \bibinfo {author} {\bibfnamefont {M.}~\bibnamefont {Tomasin}}, \bibinfo {author} {\bibfnamefont {F.}~\bibnamefont {Vedovato}}, \bibinfo {author} {\bibfnamefont {M.}~\bibnamefont {Schiavon}}, \bibinfo {author} {\bibfnamefont {V.}~\bibnamefont {Luceri}}, \bibinfo {author} {\bibfnamefont {G.}~\bibnamefont {Bianco}},\ and\ \bibinfo {author} {\bibfnamefont {P.}~\bibnamefont {Villoresi}},\ }\bibfield  {title} {\bibinfo {title} {Interference at the single photon level along satellite-ground channels},\ }\href@noop {} {\bibfield  {journal} {\bibinfo  {journal} {Physical review letters}\ }\textbf {\bibinfo {volume} {116}},\ \bibinfo {pages} {253601} (\bibinfo {year} {2016})}\BibitemShut {NoStop}%
\bibitem [{\citenamefont {Vagniluca}\ \emph {et~al.}(2020)\citenamefont {Vagniluca}, \citenamefont {Da~Lio}, \citenamefont {Rusca}, \citenamefont {Cozzolino}, \citenamefont {Ding}, \citenamefont {Zbinden}, \citenamefont {Zavatta}, \citenamefont {Oxenl{\o}we},\ and\ \citenamefont {Bacco}}]{2020_Vagniluca_TB4}%
  \BibitemOpen
  \bibfield  {author} {\bibinfo {author} {\bibfnamefont {I.}~\bibnamefont {Vagniluca}}, \bibinfo {author} {\bibfnamefont {B.}~\bibnamefont {Da~Lio}}, \bibinfo {author} {\bibfnamefont {D.}~\bibnamefont {Rusca}}, \bibinfo {author} {\bibfnamefont {D.}~\bibnamefont {Cozzolino}}, \bibinfo {author} {\bibfnamefont {Y.}~\bibnamefont {Ding}}, \bibinfo {author} {\bibfnamefont {H.}~\bibnamefont {Zbinden}}, \bibinfo {author} {\bibfnamefont {A.}~\bibnamefont {Zavatta}}, \bibinfo {author} {\bibfnamefont {L.~K.}\ \bibnamefont {Oxenl{\o}we}},\ and\ \bibinfo {author} {\bibfnamefont {D.}~\bibnamefont {Bacco}},\ }\bibfield  {title} {\bibinfo {title} {Efficient time-bin encoding for practical high-dimensional quantum key distribution},\ }\href@noop {} {\bibfield  {journal} {\bibinfo  {journal} {Physical Review Applied}\ }\textbf {\bibinfo {volume} {14}},\ \bibinfo {pages} {014051} (\bibinfo {year} {2020})}\BibitemShut {NoStop}%
\bibitem [{\citenamefont {Yu}\ \emph {et~al.}(2025)\citenamefont {Yu}, \citenamefont {Sciara}, \citenamefont {Chemnitz}, \citenamefont {Montaut}, \citenamefont {Crockett}, \citenamefont {Fischer}, \citenamefont {Helsten}, \citenamefont {Wetzel}, \citenamefont {Goebel}, \citenamefont {Kr{\"a}mer} \emph {et~al.}}]{2025_Morendotti_TB8}%
  \BibitemOpen
  \bibfield  {author} {\bibinfo {author} {\bibfnamefont {H.}~\bibnamefont {Yu}}, \bibinfo {author} {\bibfnamefont {S.}~\bibnamefont {Sciara}}, \bibinfo {author} {\bibfnamefont {M.}~\bibnamefont {Chemnitz}}, \bibinfo {author} {\bibfnamefont {N.}~\bibnamefont {Montaut}}, \bibinfo {author} {\bibfnamefont {B.}~\bibnamefont {Crockett}}, \bibinfo {author} {\bibfnamefont {B.}~\bibnamefont {Fischer}}, \bibinfo {author} {\bibfnamefont {R.}~\bibnamefont {Helsten}}, \bibinfo {author} {\bibfnamefont {B.}~\bibnamefont {Wetzel}}, \bibinfo {author} {\bibfnamefont {T.~A.}\ \bibnamefont {Goebel}}, \bibinfo {author} {\bibfnamefont {R.~G.}\ \bibnamefont {Kr{\"a}mer}}, \emph {et~al.},\ }\bibfield  {title} {\bibinfo {title} {Quantum key distribution implemented with d-level time-bin entangled photons},\ }\href@noop {} {\bibfield  {journal} {\bibinfo  {journal} {Nature Communications}\ }\textbf {\bibinfo {volume} {16}},\ \bibinfo {pages} {171} (\bibinfo {year} {2025})}\BibitemShut {NoStop}%
\bibitem [{\citenamefont {Kues}\ \emph {et~al.}(2017)\citenamefont {Kues}, \citenamefont {Reimer}, \citenamefont {Roztocki}, \citenamefont {Cort{\'e}s}, \citenamefont {Sciara}, \citenamefont {Wetzel}, \citenamefont {Zhang}, \citenamefont {Cino}, \citenamefont {Chu}, \citenamefont {Little} \emph {et~al.}}]{2017_Kues_FB}%
  \BibitemOpen
  \bibfield  {author} {\bibinfo {author} {\bibfnamefont {M.}~\bibnamefont {Kues}}, \bibinfo {author} {\bibfnamefont {C.}~\bibnamefont {Reimer}}, \bibinfo {author} {\bibfnamefont {P.}~\bibnamefont {Roztocki}}, \bibinfo {author} {\bibfnamefont {L.~R.}\ \bibnamefont {Cort{\'e}s}}, \bibinfo {author} {\bibfnamefont {S.}~\bibnamefont {Sciara}}, \bibinfo {author} {\bibfnamefont {B.}~\bibnamefont {Wetzel}}, \bibinfo {author} {\bibfnamefont {Y.}~\bibnamefont {Zhang}}, \bibinfo {author} {\bibfnamefont {A.}~\bibnamefont {Cino}}, \bibinfo {author} {\bibfnamefont {S.~T.}\ \bibnamefont {Chu}}, \bibinfo {author} {\bibfnamefont {B.~E.}\ \bibnamefont {Little}}, \emph {et~al.},\ }\bibfield  {title} {\bibinfo {title} {On-chip generation of high-dimensional entangled quantum states and their coherent control},\ }\href@noop {} {\bibfield  {journal} {\bibinfo  {journal} {Nature}\ }\textbf {\bibinfo {volume} {546}},\ \bibinfo {pages} {622} (\bibinfo {year} {2017})}\BibitemShut {NoStop}%
\bibitem [{\citenamefont {Clementi}\ \emph {et~al.}(2023)\citenamefont {Clementi}, \citenamefont {Sabattoli}, \citenamefont {Borghi}, \citenamefont {Gianini}, \citenamefont {Tagliavacche}, \citenamefont {El~Dirani}, \citenamefont {Youssef}, \citenamefont {Bergamasco}, \citenamefont {Petit-Etienne}, \citenamefont {Pargon} \emph {et~al.}}]{2023_Clementi}%
  \BibitemOpen
  \bibfield  {author} {\bibinfo {author} {\bibfnamefont {M.}~\bibnamefont {Clementi}}, \bibinfo {author} {\bibfnamefont {F.~A.}\ \bibnamefont {Sabattoli}}, \bibinfo {author} {\bibfnamefont {M.}~\bibnamefont {Borghi}}, \bibinfo {author} {\bibfnamefont {L.}~\bibnamefont {Gianini}}, \bibinfo {author} {\bibfnamefont {N.}~\bibnamefont {Tagliavacche}}, \bibinfo {author} {\bibfnamefont {H.}~\bibnamefont {El~Dirani}}, \bibinfo {author} {\bibfnamefont {L.}~\bibnamefont {Youssef}}, \bibinfo {author} {\bibfnamefont {N.}~\bibnamefont {Bergamasco}}, \bibinfo {author} {\bibfnamefont {C.}~\bibnamefont {Petit-Etienne}}, \bibinfo {author} {\bibfnamefont {E.}~\bibnamefont {Pargon}}, \emph {et~al.},\ }\bibfield  {title} {\bibinfo {title} {Programmable frequency-bin quantum states in a nano-engineered silicon device},\ }\href@noop {} {\bibfield  {journal} {\bibinfo  {journal} {Nature Communications}\ }\textbf {\bibinfo {volume} {14}},\ \bibinfo {pages} {176} (\bibinfo {year} {2023})}\BibitemShut {NoStop}%
\bibitem [{\citenamefont {Borghi}\ \emph {et~al.}(2023)\citenamefont {Borghi}, \citenamefont {Tagliavacche}, \citenamefont {Sabattoli}, \citenamefont {Dirani}, \citenamefont {Youssef}, \citenamefont {Petit-Etienne}, \citenamefont {Pargon}, \citenamefont {Sipe}, \citenamefont {Liscidini}, \citenamefont {Sciancalepore} \emph {et~al.}}]{2023_Borghi}%
  \BibitemOpen
  \bibfield  {author} {\bibinfo {author} {\bibfnamefont {M.}~\bibnamefont {Borghi}}, \bibinfo {author} {\bibfnamefont {N.}~\bibnamefont {Tagliavacche}}, \bibinfo {author} {\bibfnamefont {F.~A.}\ \bibnamefont {Sabattoli}}, \bibinfo {author} {\bibfnamefont {H.~E.}\ \bibnamefont {Dirani}}, \bibinfo {author} {\bibfnamefont {L.}~\bibnamefont {Youssef}}, \bibinfo {author} {\bibfnamefont {C.}~\bibnamefont {Petit-Etienne}}, \bibinfo {author} {\bibfnamefont {E.}~\bibnamefont {Pargon}}, \bibinfo {author} {\bibfnamefont {J.}~\bibnamefont {Sipe}}, \bibinfo {author} {\bibfnamefont {M.}~\bibnamefont {Liscidini}}, \bibinfo {author} {\bibfnamefont {C.}~\bibnamefont {Sciancalepore}}, \emph {et~al.},\ }\bibfield  {title} {\bibinfo {title} {Reconfigurable silicon photonic chip for the generation of frequency-bin-entangled qudits},\ }\href@noop {} {\bibfield  {journal} {\bibinfo  {journal} {Physical Review Applied}\ }\textbf {\bibinfo {volume} {19}},\ \bibinfo {pages} {064026} (\bibinfo {year} {2023})}\BibitemShut {NoStop}%
\bibitem [{\citenamefont {Lu}\ \emph {et~al.}(2023{\natexlab{b}})\citenamefont {Lu}, \citenamefont {Liscidini}, \citenamefont {Gaeta}, \citenamefont {Weiner},\ and\ \citenamefont {Lukens}}]{lu2023frequency}%
  \BibitemOpen
  \bibfield  {author} {\bibinfo {author} {\bibfnamefont {H.-H.}\ \bibnamefont {Lu}}, \bibinfo {author} {\bibfnamefont {M.}~\bibnamefont {Liscidini}}, \bibinfo {author} {\bibfnamefont {A.~L.}\ \bibnamefont {Gaeta}}, \bibinfo {author} {\bibfnamefont {A.~M.}\ \bibnamefont {Weiner}},\ and\ \bibinfo {author} {\bibfnamefont {J.~M.}\ \bibnamefont {Lukens}},\ }\bibfield  {title} {\bibinfo {title} {Frequency-bin photonic quantum information},\ }\href@noop {} {\bibfield  {journal} {\bibinfo  {journal} {Optica}\ }\textbf {\bibinfo {volume} {10}},\ \bibinfo {pages} {1655} (\bibinfo {year} {2023}{\natexlab{b}})}\BibitemShut {NoStop}%
\bibitem [{\citenamefont {Khodadad~Kashi}\ and\ \citenamefont {Kues}(2025)}]{2025_Kues_FB_QKD}%
  \BibitemOpen
  \bibfield  {author} {\bibinfo {author} {\bibfnamefont {A.}~\bibnamefont {Khodadad~Kashi}}\ and\ \bibinfo {author} {\bibfnamefont {M.}~\bibnamefont {Kues}},\ }\bibfield  {title} {\bibinfo {title} {Frequency-bin-encoded entanglement-based quantum key distribution in a reconfigurable frequency-multiplexed network},\ }\href@noop {} {\bibfield  {journal} {\bibinfo  {journal} {Light: Science \& Applications}\ }\textbf {\bibinfo {volume} {14}},\ \bibinfo {pages} {49} (\bibinfo {year} {2025})}\BibitemShut {NoStop}%
\bibitem [{\citenamefont {Tagliavacche}\ \emph {et~al.}(2025)\citenamefont {Tagliavacche}, \citenamefont {Borghi}, \citenamefont {Guarda}, \citenamefont {Ribezzo}, \citenamefont {Liscidini}, \citenamefont {Bacco}, \citenamefont {Galli},\ and\ \citenamefont {Bajoni}}]{2024_Tagliavacche}%
  \BibitemOpen
  \bibfield  {author} {\bibinfo {author} {\bibfnamefont {N.}~\bibnamefont {Tagliavacche}}, \bibinfo {author} {\bibfnamefont {M.}~\bibnamefont {Borghi}}, \bibinfo {author} {\bibfnamefont {G.}~\bibnamefont {Guarda}}, \bibinfo {author} {\bibfnamefont {D.}~\bibnamefont {Ribezzo}}, \bibinfo {author} {\bibfnamefont {M.}~\bibnamefont {Liscidini}}, \bibinfo {author} {\bibfnamefont {D.}~\bibnamefont {Bacco}}, \bibinfo {author} {\bibfnamefont {M.}~\bibnamefont {Galli}},\ and\ \bibinfo {author} {\bibfnamefont {D.}~\bibnamefont {Bajoni}},\ }\bibfield  {title} {\bibinfo {title} {Frequency-bin entanglement-based quantum key distribution},\ }\href@noop {} {\bibfield  {journal} {\bibinfo  {journal} {npj Quantum Information}\ }\textbf {\bibinfo {volume} {11}},\ \bibinfo {pages} {60} (\bibinfo {year} {2025})}\BibitemShut {NoStop}%
\bibitem [{\citenamefont {Henry}\ \emph {et~al.}(2023)\citenamefont {Henry}, \citenamefont {Raghunathan}, \citenamefont {Ricard}, \citenamefont {Lefaucher}, \citenamefont {Miatto}, \citenamefont {Belabas}, \citenamefont {Zaquine},\ and\ \citenamefont {All{\'e}aume}}]{henry2023parallelizable}%
  \BibitemOpen
  \bibfield  {author} {\bibinfo {author} {\bibfnamefont {A.}~\bibnamefont {Henry}}, \bibinfo {author} {\bibfnamefont {R.}~\bibnamefont {Raghunathan}}, \bibinfo {author} {\bibfnamefont {G.}~\bibnamefont {Ricard}}, \bibinfo {author} {\bibfnamefont {B.}~\bibnamefont {Lefaucher}}, \bibinfo {author} {\bibfnamefont {F.}~\bibnamefont {Miatto}}, \bibinfo {author} {\bibfnamefont {N.}~\bibnamefont {Belabas}}, \bibinfo {author} {\bibfnamefont {I.}~\bibnamefont {Zaquine}},\ and\ \bibinfo {author} {\bibfnamefont {R.}~\bibnamefont {All{\'e}aume}},\ }\bibfield  {title} {\bibinfo {title} {Parallelizable synthesis of arbitrary single-qubit gates with linear optics and time-frequency encoding},\ }\href@noop {} {\bibfield  {journal} {\bibinfo  {journal} {Physical Review A}\ }\textbf {\bibinfo {volume} {107}},\ \bibinfo {pages} {062610} (\bibinfo {year} {2023})}\BibitemShut {NoStop}%
\bibitem [{\citenamefont {Liscidini}\ and\ \citenamefont {Sipe}(2019)}]{2019_Liscidini_FBtheory}%
  \BibitemOpen
  \bibfield  {author} {\bibinfo {author} {\bibfnamefont {M.}~\bibnamefont {Liscidini}}\ and\ \bibinfo {author} {\bibfnamefont {J.}~\bibnamefont {Sipe}},\ }\bibfield  {title} {\bibinfo {title} {Scalable and efficient source of entangled frequency bins},\ }\href@noop {} {\bibfield  {journal} {\bibinfo  {journal} {Optics Letters}\ }\textbf {\bibinfo {volume} {44}},\ \bibinfo {pages} {2625} (\bibinfo {year} {2019})}\BibitemShut {NoStop}%
\bibitem [{\citenamefont {Vallone}\ \emph {et~al.}(2008{\natexlab{b}})\citenamefont {Vallone}, \citenamefont {Ceccarelli}, \citenamefont {De~Martini},\ and\ \citenamefont {Mataloni}}]{2008_Vallone_witness}%
  \BibitemOpen
  \bibfield  {author} {\bibinfo {author} {\bibfnamefont {G.}~\bibnamefont {Vallone}}, \bibinfo {author} {\bibfnamefont {R.}~\bibnamefont {Ceccarelli}}, \bibinfo {author} {\bibfnamefont {F.}~\bibnamefont {De~Martini}},\ and\ \bibinfo {author} {\bibfnamefont {P.}~\bibnamefont {Mataloni}},\ }\bibfield  {title} {\bibinfo {title} {Hyperentanglement witness},\ }\href@noop {} {\bibfield  {journal} {\bibinfo  {journal} {Physical Review A-Atomic, Molecular, and Optical Physics}\ }\textbf {\bibinfo {volume} {78}},\ \bibinfo {pages} {062305} (\bibinfo {year} {2008}{\natexlab{b}})}\BibitemShut {NoStop}%
\bibitem [{\citenamefont {Borghi}\ \emph {et~al.}(2025)\citenamefont {Borghi}, \citenamefont {Brusaschi}, \citenamefont {Liscidini}, \citenamefont {Galli},\ and\ \citenamefont {Bajoni}}]{2025_Borghi}%
  \BibitemOpen
  \bibfield  {author} {\bibinfo {author} {\bibfnamefont {M.}~\bibnamefont {Borghi}}, \bibinfo {author} {\bibfnamefont {E.}~\bibnamefont {Brusaschi}}, \bibinfo {author} {\bibfnamefont {M.}~\bibnamefont {Liscidini}}, \bibinfo {author} {\bibfnamefont {M.}~\bibnamefont {Galli}},\ and\ \bibinfo {author} {\bibfnamefont {D.}~\bibnamefont {Bajoni}},\ }\bibfield  {title} {\bibinfo {title} {Bipartite Gaussian boson sampling in the time-frequency-bin domain with squeezed light generated by a silicon nitride microresonator},\ }\href@noop {} {\bibfield  {journal} {\bibinfo  {journal} {npj Quantum Information}\ } \textbf {\bibinfo {volume} {11}},\ \bibinfo {pages} {135} (\bibinfo {year} {2025})}\BibitemShut {NoStop}%
\bibitem [{\citenamefont {Takesue}\ and\ \citenamefont {Noguchi}(2009)}]{takesue2009implementation}%
  \BibitemOpen
  \bibfield  {author} {\bibinfo {author} {\bibfnamefont {H.}~\bibnamefont {Takesue}}\ and\ \bibinfo {author} {\bibfnamefont {Y.}~\bibnamefont {Noguchi}},\ }\bibfield  {title} {\bibinfo {title} {Implementation of quantum state tomography for time-bin entangled photon pairs},\ }\href@noop {} {\bibfield  {journal} {\bibinfo  {journal} {Optics express}\ }\textbf {\bibinfo {volume} {17}},\ \bibinfo {pages} {10976} (\bibinfo {year} {2009})}\BibitemShut {NoStop}%
\end{thebibliography}
%%% for Arxiv:
%%%\bibliographystyle{apsrev4-2}
%%%\input{output.bbl}  

%apsrev4-2.bst 2019-01-14 (MD) hand-edited version of apsrev4-1.bst
%Control: key (0)
%Control: author (8) initials jnrlst
%Control: editor formatted (1) identically to author
%Control: production of article title (0) allowed
%Control: page (0) single
%Control: year (1) truncated
%Control: production of eprint (0) enabled
%

\end{document}